\documentclass[aps,prl,preprint,superscriptaddress]{revtex4}

\usepackage{graphicx,amsmath,amssymb}

\begin{document}

%\preprint{}

\title{Interaction studies of a heavy-light meson-baryon system}
\thanks{This material is based upon work supported by the National Science Foundation
under Grant No. 0300065 and upon resources provided by the Lattice Hadron Physics
Collaboration LHPC through the SciDac program of the US Department of Energy.}
\author{M.S. Cook}
\author{H.R. Fiebig}
\affiliation{Physics Department, FIU-University Park, Miami, Florida 33199, USA}
\collaboration{LHP Collaboration}

\date{\today}

\begin{abstract}
We study time correlation functions of operators representing
heavy-light $K$--$\Lambda$ like systems at various relative
distances $r$. The heavy quarks, one in each hadron, are
treated as static.  An anisotropic and asymmetric lattice is
used with Wilson fermions. Our goal is to extract an adiabatic
potential and thus learn about the physics of the five-quark
system viewed as an hadronic molecule.
\end{abstract}


\maketitle

\section{Introduction}

Lattice QCD studies of hadron-hadron interactions are the gateway to
nuclear physics through first principles \cite{Savage:2005,Fiebig:2002kg}. 
From a lattice simulation point of view the nucleon-nucleon interaction is
probably the most challenging case. This is evident given the large spatial
size of the deuteron, and in particular, the insight that the physics of the
long-range strong interaction is driven mostly by the pion
cloud \cite{Machleidt:1989tm}. The latter may be taken as an indication that
chiral symmetry and a full (unquenched) lattice action are high priorities for simulations
aiming at quantitative results.

Aside from the above prominent case, however, interactions in other two-hadron systems are worth
investigating as well, because this might lead to new insights into the structural features of
some of the experimentally known baryon resonances \cite{PDBook}.
In particular, we here ask if some of those may be understood as hadronic molecules, similar
to the deuteron, but possibly with different physics of the interaction mechanisms in which
quark and gluon degrees of freedom play a role.
Prime candidates for such systems are pairs of hadrons containing one heavy quark each because,
in the spirit of the Born-Oppenheimer approximation, the (slow) heavy quarks naturally
serve to define the centers of two hadrons while the (fast) light quarks and gluons
provide the physics of the interaction.
Studies along those lines have been done before in the context of meson-meson and
baryon-baryon systems
\cite{Arndt:2003vx,Michael:1999nq,Mihaly:1997ue}.

We here report on the current status of interaction studies of a heavy-light
meson-baryon (five-quark) hadron with the quantum numbers of an s-wave $K$--$\Lambda$
system. Because the static approximation is employed for the heavy-quark propagator,
the total energy can be computed as a function of the relative
distance $r$ between the heavy quarks. Thus an adiabatic (Born-Oppenheimer) potential
$V_a(r)$ is extracted. This can be used to address the possibility of molecule-like
structures.

\section{Simulation details}

\begin{figure}
\noindent\rule{1mm}{0mm}
\includegraphics[angle=0,width=27mm]{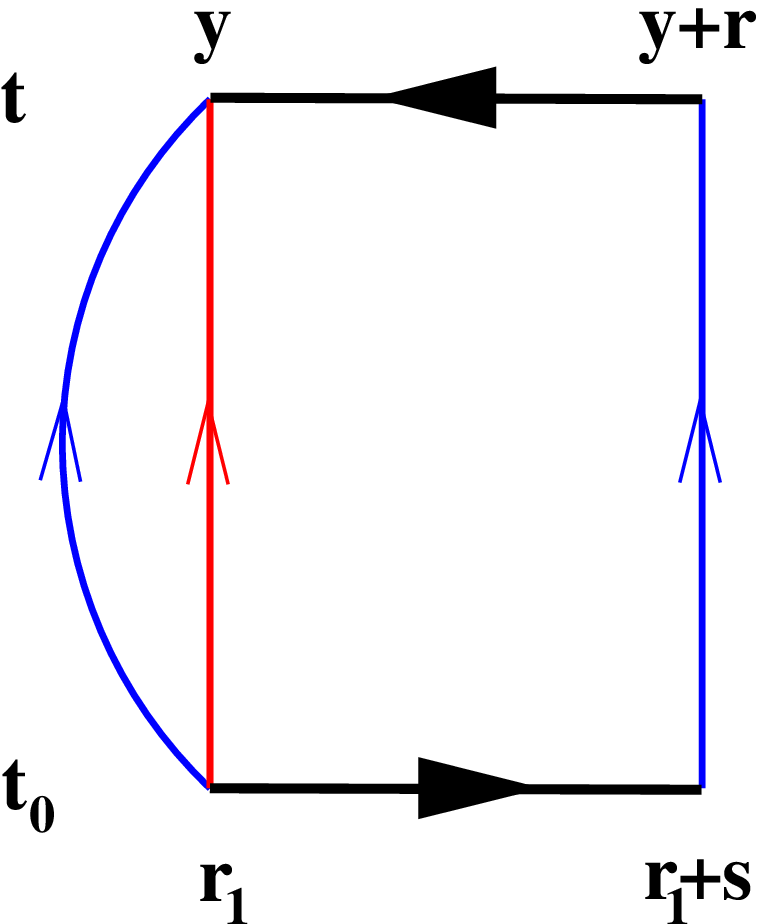}\hspace{15mm}
\includegraphics[angle=0,width=27mm]{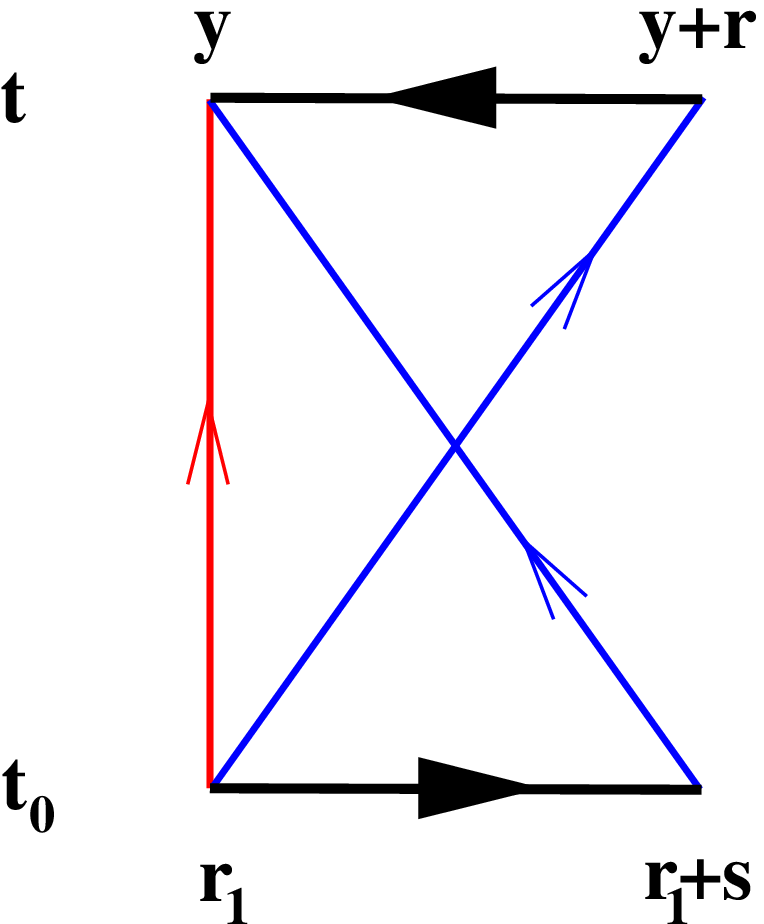}\hspace{15mm}
\includegraphics[angle=0,width=27mm]{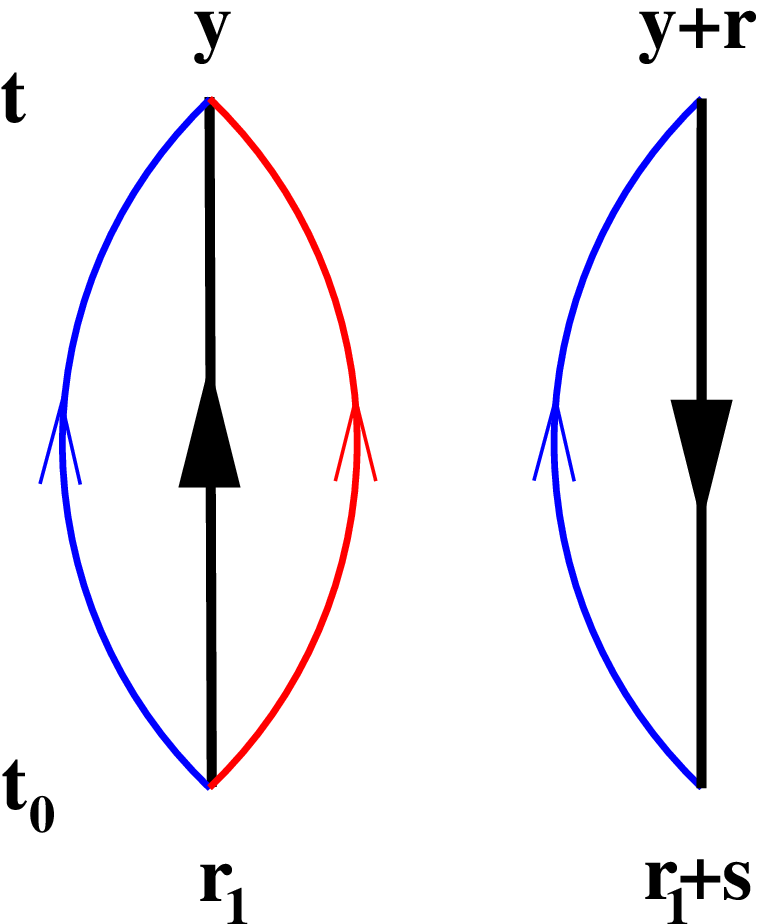}\hspace{15mm}
\includegraphics[angle=0,width=27mm]{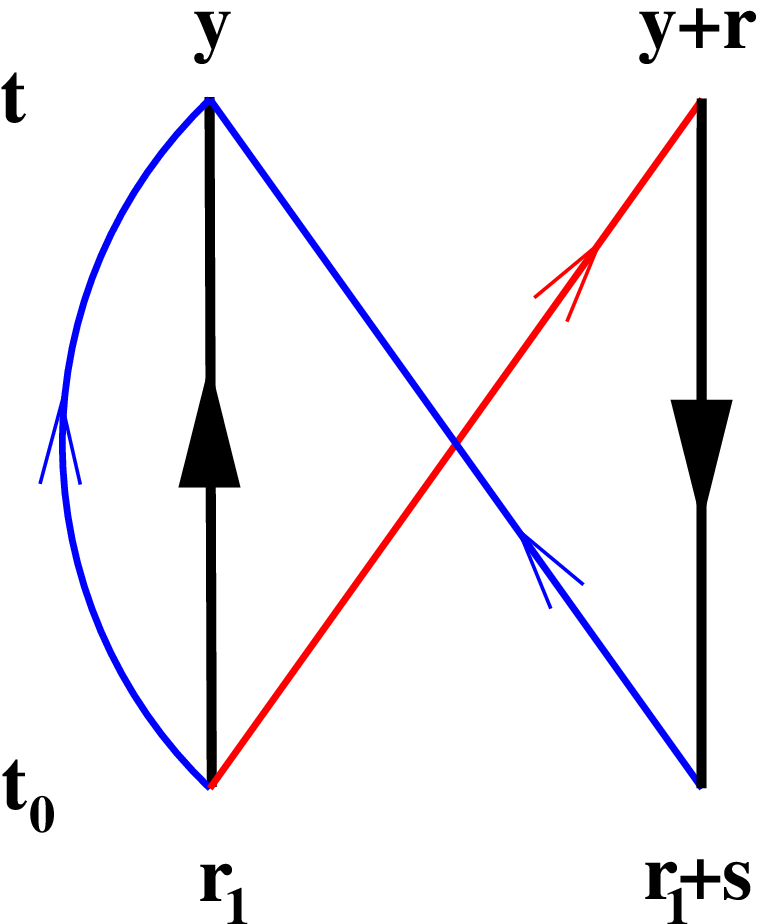}\rule{2mm}{0mm}
\caption{Diagrammatic representation of (\protect\ref{CHG}).
Thick (black) lines indicate heavy-quark propagators, and thin (blue and red)
lines depict light (u and d) quark propagators.}
\label{fig1}\end{figure}

Two-hadron interpolating fields are constructed from standard
local operators for the $K^+$ and the $\Lambda^0$ particles \cite{Mon94}
at relative distance $\vec{r}$ and projected to total momentum zero
\begin{equation}\label{Op2}
{\cal O}_\alpha(\vec{r};t)=V^{-1/2}\sum_{\vec{x}}\sum_{\vec{y}}
\delta_{\textstyle\vec{r},\vec{x}-\vec{y}}\,
K^+(\vec{x}t)\Lambda^0_\alpha(\vec{y}t)\,.
\end{equation}
Here $V$ is the spatial lattice volume and $\alpha$ is a Dirac spinor index.
Then, with
\begin{equation}
\bar{\cal O}_\mu(\vec{r};t)={\cal O}^\dagger_\alpha(\vec{r};t)\gamma_{4,\alpha\mu}
\end{equation}
the correlation function
\begin{equation}
C= \langle {\cal O}_\mu(\vec{r};t) \bar{\cal O}_\mu(\vec{s};t_0) \rangle
-\langle {\cal O}_\mu(\vec{r};t) \rangle\langle \bar{\cal O}_\mu(\vec{s};t_0) \rangle\,,
\end{equation}
where $\vec{s}$ is the relative distance at the source,
can be expressed in terms of fermion propagators. The flavor assignment
$K^+\Lambda^0\sim \bar{s}u\,uds$ causes the separable term to vanish.
Writing $H(\vec{x}t,\vec{y}t_0)$ and $G(\vec{x}t,\vec{y}t_0)$
for the heavy (s) and light (u,d) quark propagators, respectively, one obtains
\begin{eqnarray}\label{CHG}
C&=&\langle\sum_{\vec{y}}
[H(\vec{y}t,\vec{y}+\vec{r}t)H(\vec{r}_1+\vec{s}t_0,\vec{r}_1t_0)
-H(\vec{y}t,\vec{r}_1t_0)H(\vec{r}_1+\vec{s}t_0,\vec{y}+\vec{r}t)]\\
&&\rule{3.7ex}{0ex}
G(\vec{y}t,\vec{r}_1t_0)
[G(\vec{y}t,\vec{r}_1t_0)G(\vec{y}+\vec{r}_1t,\vec{r}_1+\vec{s}t_0)
-G(\vec{y}t,\vec{r}_1+\vec{s}t_0)G(\vec{y}+\vec{r}_1t,\vec{r}_1t_0)]\rangle\,.
\nonumber
\end{eqnarray}
For clarity the rather involved color and spin index structure is not shown in (\ref{CHG}).
Also, translational invariance has been used to arrive at the above expression, and an
arbitrary space site $\vec{r}_1$ was introduced in this context.
A diagrammatic representation of (\ref{CHG}) is shown in Fig.~\ref{fig1}.

The heavy-quark propagators are employed in the static approximation.
For (unimproved) Wilson fermions with hopping parameter $\kappa$ this
means that the propagator is taken in the limit $\kappa\rightarrow0$ ,
resulting in
\begin{equation}\label{Hstatic}
H(\vec{x}t,\vec{y}t_0) = \delta_{\textstyle \vec{x},\vec{y}}\;(2\kappa)^{t-t_0}
\frac{1}{2}(1+\gamma_4)\,\mathcal{U}^\dagger(\vec{x};t_0t)\,,
\end{equation}
where $\mathcal{U}(\vec{x};t_0t)$ is the product of $SU(3)$ link variables
along a straight line from $(\vec{x}t_0)$ to $(\vec{x}t)$ \cite{Mon94}. 

The distance $\vec{r}=0$ is rather special \cite{Michael:1999nq} because a color singlet
operator, as realized by (\ref{Op2}), can also be achieved by a ``color twisted'' version
of (\ref{Op2}) where quarks in $K^+$ and $\Lambda^0$ are combined into a color singlet.
Because we do not consider color twisted operators in this work we restrict ourselves
to non-zero relative distance.
Thus, because $H(\vec{y}t,\vec{y}+\vec{r}t)\propto \delta_{\vec{r},\vec{0}}$
and $H(\vec{r}_1+\vec{s}t_0,\vec{r}_1t_0)\propto \delta_{\vec{s},\vec{0}}$,
only the last two diagrams in Fig.~\ref{fig1} make a contribution to
the correlation function for non-zero relative distance.
By way of (\ref{Hstatic}) those diagrams in Fig.~\ref{fig1} are proportional to
$\delta_{\vec{y},\vec{r}_1}\delta_{\vec{r},\vec{s}}$. Thus (\ref{CHG}) becomes
\begin{eqnarray}\label{C}
C&=&\delta_{\textstyle\vec{r},\vec{r}_2-\vec{r}_1}
\;\langle H(\vec{r}_1t,\vec{r}_1t_0)H(\vec{r}_2t_0,\vec{r}_2t)\\
&&G(\vec{r}_1t,\vec{r}_1t_0)
[-G(\vec{r}_1t,\vec{r}_1t_0)G(\vec{r}_2t,\vec{r}_2t_0)
+G(\vec{r}_1t,\vec{r}_2t_0)G(\vec{r}_2t,\vec{r}_1t_0)]\rangle
\quad\mbox{for}\quad\vec{r}\neq 0\nonumber\,,
\end{eqnarray}
where $\vec{r}_2$ is yet another arbitrary space site.
We observe that sources at fixed spatial sites only are needed.
However, an undesirable consequence of the static approximation is that
the site sum $\sum_{\vec{y}}$ has vanished from (\ref{CHG}) and thus is no
longer working to improve statistics. 

The final correlator we use is extended from (\ref{C}) to a matrix by
employing several levels $k=1\ldots K$ of operator smearing. The procedure
amounts to replacing in (\ref{Op2}) all light-quark fields $\psi,\bar{\psi}$
with smeared fields $\psi^{\{k\}},\bar{\psi}^{\{k\}}$.
The computation of light-quark propagators thus requires
various levels of smearing at the source and
at the sink. We have used APE-style gauge field fuzzing \cite{Alb87a}
and Wuppertal fermion smearing \cite{Alexandrou:1994ti} with
common values for the strength parameters $\alpha=2.5$ and the number $k$
of iterations. No smearing, nor link variable fuzzing, was done for the heavy,
static, quark fields in order to preserve spatial locality,
i.e. the $\delta$ factor in (\ref{Hstatic}).
Thus, writing
$\mathcal{O} \rightarrow \mathcal{O}^{\{k\}} =
\mathcal{O}[\psi^{\{k\}},\bar{\psi}^{\{k\}}\ldots]$,
the correlator (\ref{C}) becomes a $K\times K$ matrix
\begin{equation}\label{CKK}
C^{\,k\ell}(\vec{r};t,t_0)=
\langle\mathcal{O}{}^{\{k\}}_{\mu}(\vec{r};t)\,
\bar{\mathcal{O}}{}^{\,\{\ell\}}_{\mu}(\vec{r};t_0)\rangle
\quad\mbox{with}\quad k,\ell=1\cdots K\,.
\end{equation}
The expression for $C^{\,k\ell}$ in terms of quark propagators still
has the form given by (\ref{C}), however, light propagator elements
are replaced with smeared ones, $G\rightarrow G^{\,k\ell}$,
with appropriate smearing levels at source and sink.
The correlation matrix (\ref{CKK}) is hermitian by construction.

The lattice geometry is chosen as
$L_1\times L_2\times L_3\times L_4 = 8\times 8\times 32\times 16$
with bare lattice constants $a_1=a_2=2a_3=2a_4$ in the respective directions.
This choice of an asymmetric and anisotropic lattice provides a fine mass
resolution, in t-direction, and the same spatial resolution for the
adiabatic potential, as the static sources are placed along the z-direction.
The positions of the latter are at
$x=(5,5,n,3)$ with $n=1,2,3,4,8,11,13,17$, with $\vec{r}_{1,2}=(5,5,n)$ in
reference to (\ref{C}). In this way all possible relative distances
$r=|\vec{r}|$ in the range $r=1\ldots 16$ can be obtained. 
Note that periodic boundary conditions, in the spatial directions, allow us to only utilize
half of the extent of the lattice in z-direction. 

We have used the Wilson plaquette action with Wilson fermions in a quenched simulation.
The gauge field couplings in the $\mu$-$\nu$ planes and the hopping parameters in directions
$\mu$ are given by, respectively,
\begin{equation}
\beta_{\mu\nu}=\beta\,\frac{a_1a_2a_3a_4}{(a_{\mu}a_{\nu})^2}
\quad\mbox{and}\quad
\kappa_\mu = \frac{\kappa}{a_\mu\frac{1}{4}\sum_{\nu=1}^{4}\frac{1}{a_\nu}}\,.
\end{equation}
The simulation was done at $\beta=6.2$ with four values 
$\kappa=0.140,\, 0.136,\, 0.132,\, 0.128$ of hopping parameters using a multiple mass
solver \cite{Glassner:1996gz}.

\section{Analysis}

The correlation matrix (\ref{CKK}) is constructed with three smearing levels, $K=3$.
It is then diagonalized separately on each timeslice $t$, using singular value decomposition.
For asymptotic $t$ the largest eigenvalues correspond to the
ground state of the two-hadron system. The maximum entropy method (MEM),
as implemented in \cite{Fiebig:2002sp}, is used to extract masses.

In order to set the physical mass scale the nucleon (N) and vector meson ($\rho$) masses
are fit to the model $y=c_1+c_2x+c_3\ln(1+x)$, with $x=(am_\pi)^2$, using the four
available hopping parameters.
Here $a=a_3=a_4$ is the common lattice constant in the $z$ and $t$ directions.
The extrapolated values of $am_N$ and $am_\rho$, as
$x\rightarrow 0$, are used to set the reduced mass of the $N$--$\rho$ system to its
experimental value $m_{N\rho}=424$MeV.
This gives $a=0.096$fm ($a^{-1}=2055$MeV) .

\begin{figure}
\hspace{-1mm}
\includegraphics[angle=90,width=77mm]{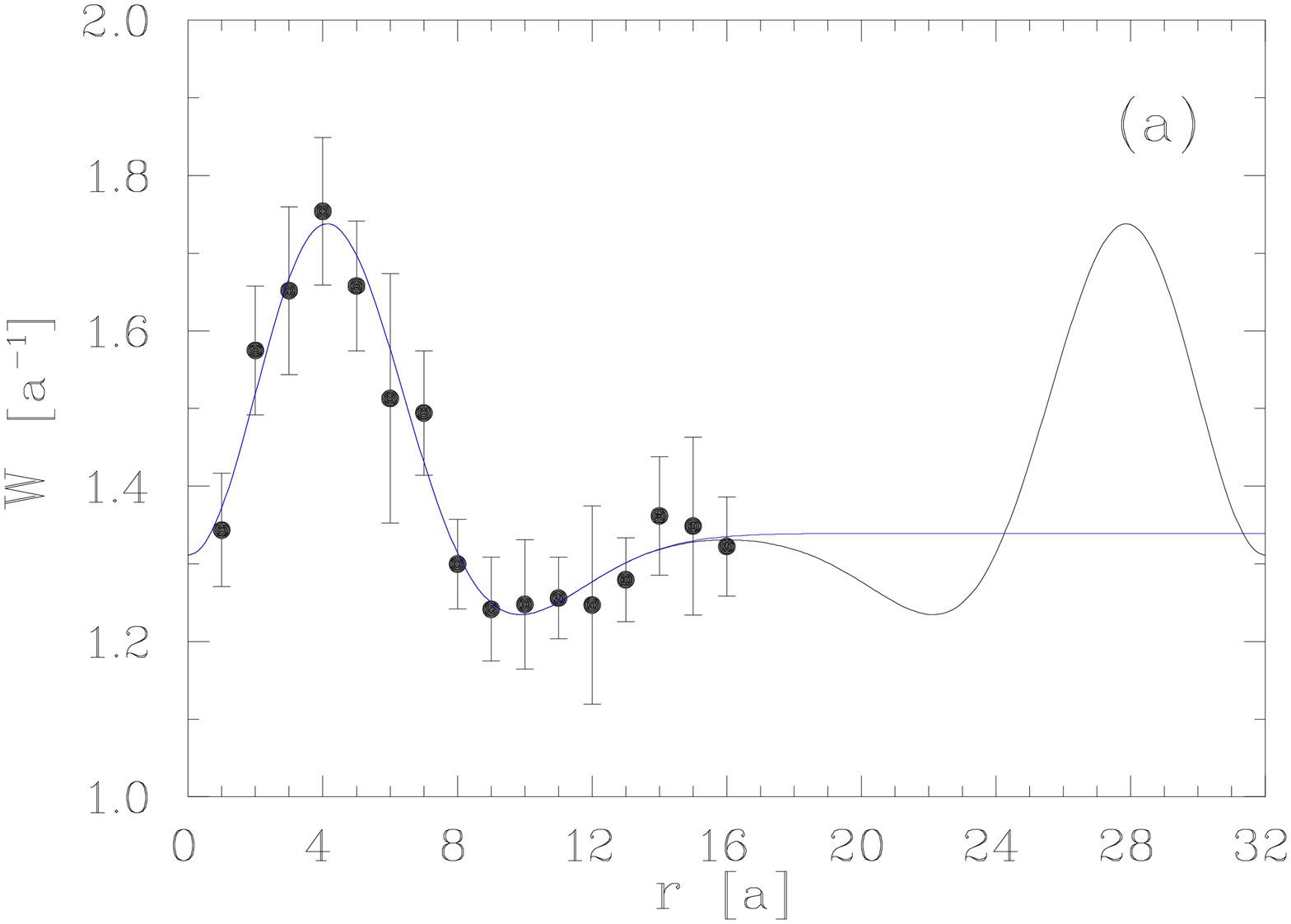}\hspace{6mm}
\includegraphics[angle=90,width=79mm]{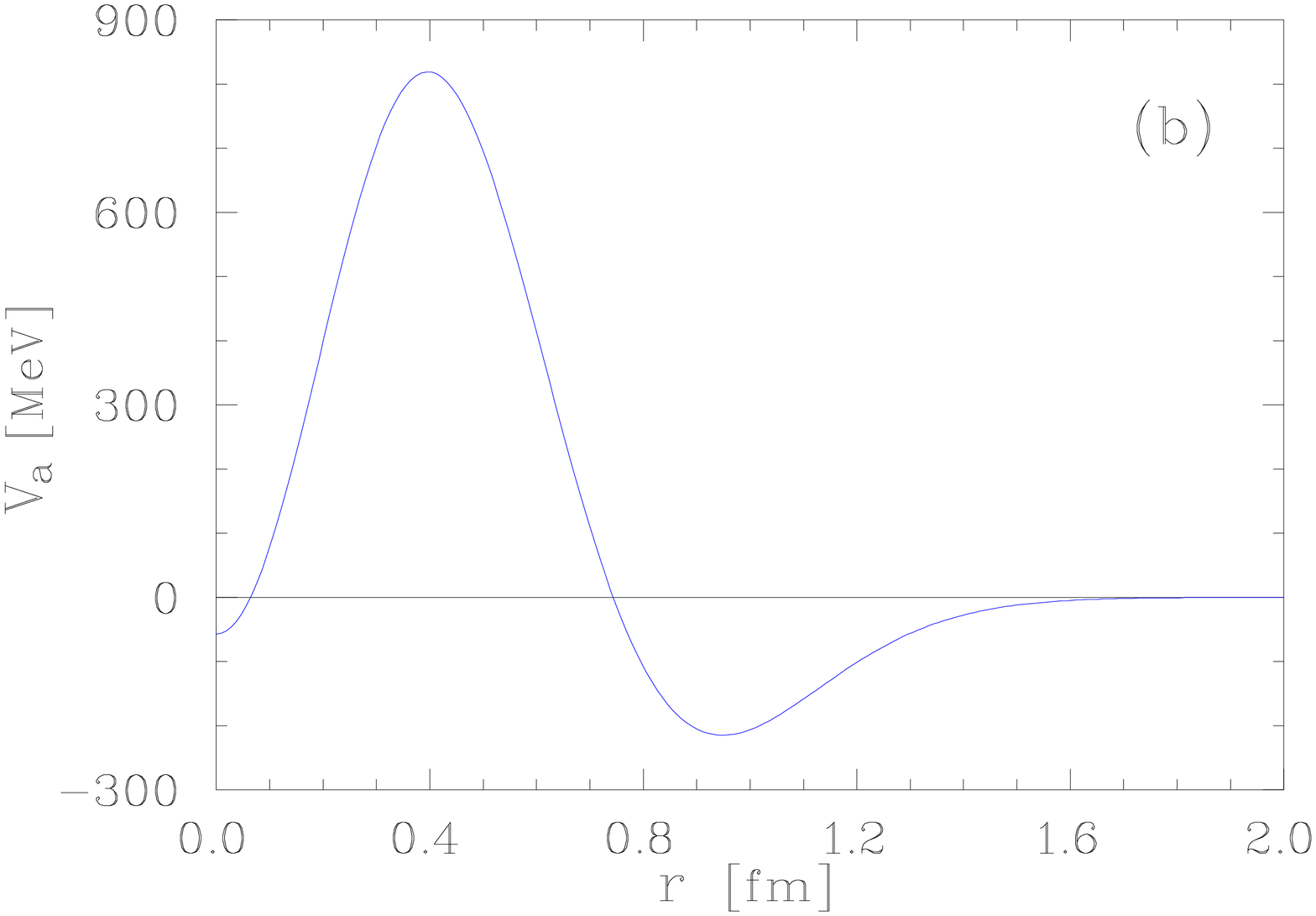}
\caption{Total energy $W$ of the heavy-light $K$-$\Lambda$ like system versus the relative
distance $r$ with MEM based uncertainties (a). The continuous lines represent fits
with (\protect\ref{V}) and (\protect\ref{VL}). The adiabatic potential $V_a$ is shown in
physical units (b). Results are for $\kappa=0.140$.}
\label{fig3}\end{figure}

At the time of this writing the analysis is limited to 90 gauge configurations
and to the largest value of the hopping parameter, $\kappa=0.140$.
The corresponding results for the total energy $W$ are shown in Fig.~\ref{fig3}a.
The uncertainties are the (rms) widths of the spectral peaks emerging from the MEM analysis.
Due to periodic boundary conditions the data points replicate for $r>16a$.
For a preliminary analysis we use the simple model 
\begin{equation}\label{V}
aV(x)=\exp(-\alpha_1x^2)(\alpha_2+\alpha_3x^2+\alpha_4x^4)\quad\mbox{with}\quad x=r/a\,.
\end{equation}
The fit to the data is then made using the periodic replication 
\begin{equation}\label{VL}
aV_L(x)=\alpha_0+aV(x)+aV(L_3-x),\quad L_3=32\,,
\end{equation}
of the model, which then has five parameters $\alpha_0\ldots\alpha_4$.
Figure~\ref{fig3}a shows the corresponding result. The plot
in Fig.~\ref{fig3}b represents (\ref{V}) in terms of the physical scale,
the adiabatic potential is $V_a(r)=V(r/a)$. 
Pending an anticipated increase of gauge configurations no error analysis for $V$
has been done
at this time. However, error bands may be inferred from the uncertainties on the data
points in Fig.~\ref{fig3}a.

\begin{figure}
\center
\includegraphics[angle=90,width=82mm]{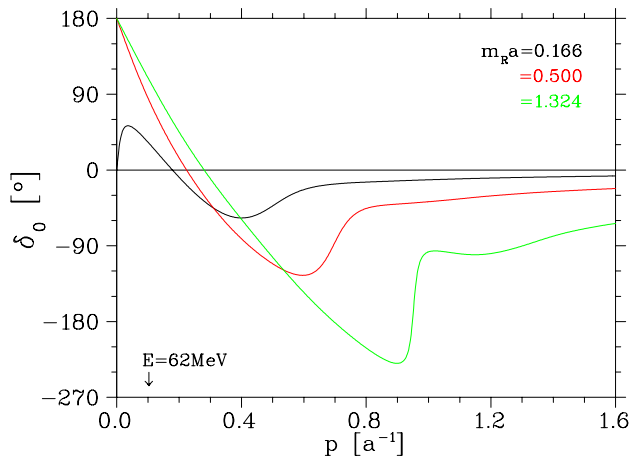}
\caption{Scattering phase shifts (s-wave) for the $K$--$\Lambda$, $D$--$\Lambda_c$,
and $B$--$\Lambda_b$ systems (black, red, and green, respectively) versus the, nonrelativistic,
relative momentum $p$. The arrow indicates the relative kinetic energy for the $K$--$\Lambda$
system at $p=0.1a^{-1}$.}
\label{fig4}\end{figure}

The attractive dip of $V_a(r)$ at around $r\approx 0.8$--$1.2$fm, in Fig.~\ref{fig3}b,
may be significant enough to produce a molecule-like structure. As a first
attempt to calculate phase shifts, we solve a
standard non-relativistic scattering problem (Schr{\"o}dinger equation) employing the
computed adiabatic potential $V_a$ with several values $m_R$ for the reduced mass.
For the latter we use the (experimental) reduced masses of the systems
$K$--$\Lambda$, $D$--$\Lambda_c$, and $B$--$\Lambda_b$ from \cite{PDBook}.
The resulting s-wave scattering phase shifts $\delta_0(p)$ are shown in Fig.~\ref{fig4}.
According to Levinson's theorem ($\delta_\ell(0)-\delta_\ell(\infty)=n\pi$) the
number $n$ of bound states is zero for the $K$--$\Lambda$ system. However, there is indication
of an emerging resonance (rising phase shift) somewhat below $p=0.1a^{-1}$, or $E=62$MeV in terms
of the relative kinetic energy. It should be kept in mind that Fig.~\ref{fig4} reflects a
result at $m_\pi>0$ ($\kappa=0.140$). At present the strength of this feature as $m_\pi\rightarrow 0$
is an unresolved question. Also, relativistic effects are not taken into account.
The latter are less significant for the $D$--$\Lambda_c$ and $B$--$\Lambda_b$ systems, respectively.
Those appear to be bound by the adiabatic potential.

\section{Assessment}

Although in its present state the simulation is not conclusive, the results give a
hint at potentially interesting physics. It appears conceivable that the known hadron mass spectrum
may contain five-quark hadrons with a molecule-like structure.  
Our preliminary results would point to a resonant state with an excitation energy typical of a nuclear
system, say $\approx 50\pm 50$MeV. As a possible candidate with the appropriate quantum numbers
the N(1650) comes to mind \cite{PDBook}. Its mass lies just 40MeV above the $K$--$\Lambda$ threshold.
However, we should caution that the extraction of masses typical for nuclear physics
from a lattice simulation is difficult, because residual hadron-hadron interactions are,
at least, one order of magnitude less that baryon rest masses. 
Therefore, awaiting the analysis with a reasonable number of gauge configurations, the above
scenario should be considered an interesting possibility.

\end{document}